\documentclass[letterpaper]{JHEP3}
\usepackage{epsfig}

\title{Decay of $C/Z_n$: exact supergravity solutions}

\author{
Matthew Headrick \\ 
MIT Center for Theoretical Physics, 77 Massachusetts Ave, Cambridge MA 02139, USA \\
E-mail: \email{headrick@mit.edu}
}

\abstract{We present a simple method to derive the general exact solution describing monopole scalar radiation coupled to gravity in $2+1$ dimensions. The solution confirms the conjecture of Adams, Polchinski, and Silverstein regarding the late time behavior of the decay of the $C/Z_n$ orbifold in type II string theory.}


\preprint{\hepth{0312213}\\MIT-CTP-3458}

\begin{document}

\section{Introduction}

Despite the impressive progress that has been made in the last several years in our understanding of open string tachyon condensation, the problem of closed string tachyon condensation remains deep and mysterious. Among the many advantages enjoyed by those studying open string tachyons,\footnote{See \cite{Martinec} for a nice review of open and closed string tachyon condensation, emphasizing the differences between the two cases.} perhaps the most fundamental is that, by taking the string coupling small, they can decouple the closed string degrees of freedom and work in a fixed gravitational background. Obviously, when studying closed string tachyons no such consistent truncation is possible, and one is forced to deal with the often complicated gravitational dynamics as an essential part of the condensation process. For example, since the energy density released during closed string tachyon condensation is proportional to $1/G_{\rm N}$ (compared to $1/\sqrt{G_{\rm N}}$ in the open string case), the gravitational back-reaction can never be neglected.

The situation is actually worse than this. Given that, generically, a string scale (times the above factor of $1/G_{\rm N}$) energy density is being released, one would expect that massive string modes get excited during the decay, and that a fully stringy treatment is therefore necessary. There is no guarantee that the notion of a smooth geometry described by the massless fields continues to be valid; most dramatically, the spacetime itself could cataclysmically decay into nothing. However, if the tachyon locus has a transverse dimension for the energy to dissipate into (and if the decay doesn't somehow open up a tear in spacetime), then it is plausible that at late times the massive string fields will dump their energy into the massless ones. In that case, as time goes on stringy effects will become less and less important, and eventually one will end up with a conventional gravity scenario in which energy propagates away from the tachyon locus at the speed of light, with a concomitant back-reaction of the geometry. This is essentially the picture proposed by Adams, Polchinski, and Silverstein \cite{Adams} in their study of the decay of the $C/Z_n$ orbifold of type II string theory: within a few string times,\footnote{H. Liu has argued that this process could in fact take a time inversely proportional to the string coupling.} the energy released in the decay is converted into a circular dilaton wave that expands at the speed of light, leaving behind a bubble of flat space.

Typically in general relativity, even such ``pedestrian" back-reaction problems are difficult or intractable, and one often resorts to an approximation in which the back-reaction of the geometry is treated as a perturbation on the equations of motion of the matter fields. The purpose of this note is to point out that, in the example of $C/Z_n$, there exists a simple method for solving the coupled dilaton-graviton equations of motion exactly and generally. Thus the back-reaction problem need not be an obstacle to understanding the decay process. In particular, the solution confirms the conjecture of APS regarding the late-time evolution of the system.

It is an oft-heard speculation that tachyon condensation processes are modelled by an RG flow of the worldsheet theory seeded by the tachyon vertex operator (which is relevant when stripped of its time dependence) \cite{Martinec}--\cite{Minwalla}. How the word ``modelled" should be precisely understood in this context depends on who is speaking, and to understand the connection better it may be useful to have examples of decay processes in which both the time evolution and the RG flow are known explicitly. An exact solution to the RG equations for this system is already known \cite{Gutperle} (this solution is valid only when the curvature scales are much larger than the string scale, just as for the time evolution solution considered in this paper). Thus, $C/Z_n$ can in principle now serve as a laboratory for testing ideas about the quantitative connection between these two types of evolution.

In Section 2 below, we set up the equations of motion for the system, namely a massless scalar minimally coupled to 2+1 gravity. We then show, using the rotational symmetry and making a judicious choice of coordinate system, that the equation of motion for the dilaton is the same as that for a dilaton in Minkowski space, i.e.\ is decoupled from the back-reaction of the geometry. This allows the equations to be solved in two steps: first, the dilaton equation of motion is straightforwardly solved, then the back-reaction of the geometry is found by solving a first-order equation for a metric coefficient that is sourced by the dilaton's energy. Among other things, this method guarantees that under non-singular initial conditions the energy will indeed radiate to infinity and that no singularity can form.

In Section 3, we study solutions to the equations of motion. The quantitative behavior of the system at late times for generic initial conditions is easily found. One general result is that in the final state the dilaton returns to its initial value---the tachyon condensation does not lead the system to run either to strong or to weak coupling. A numerical example is presented to illustrate quantitatively how the dilaton wave carries the spatial curvature with it, leaving behind a bubble of flat space. In the limit of a spatially localized dilaton source, a formula is derived for the change in the deficit angle in terms of the source's frequency spectrum. Finally, a thin-shell solution is derived for the limit in which the source is localized both in the space and in time.

This problem has previously been solved in the strong coupling limit by Gregory and Harvey \cite{Gregory}. Their method is somewhat more involved than ours but their results are consistent with ours.

\section{Equations of motion}

We wish to study the decay of the unstable $C/Z_n$ orbifold of type II string theory as a classical time-dependent process. We can consistently truncate the dynamics to $2+1$ dimensions, namely the two of the orbifold plus time \cite{Adams}.\footnote{The reason is that the CFT describing the extra seven dimensions is coupled to the $2+1$ dimensional CFT neither in the initial configuration nor by the tachyon vertex operator, and therefore decouples entirely. In other words, the on-shell tachyon vertex operator must extend to an exactly marginal operator in the $2+1$ dimensional CFT alone.} The extra seven dimensions may be in the form of $R^7$, or any other unitary $\hat c=7$ CFT. The equation of motion for the $B$-field requires its field strength to be a constant multiple of the volume form on the unreduced $2+1$ dimensions. This constant vanishes in the initial orbifold configuration, and must
therefore vanish everywhere. So the only active NS-NS fields are the metric and the dilaton $\Phi$, and their action is
\begin{equation}
S = \frac1{2\kappa_3^2}\int d^3\!x\sqrt{-G}e^{-2\Phi}
\left( R_3 + 4G^{\mu\nu}\partial_\mu\Phi\partial_\nu\Phi\right),
\end{equation}
The analysis is simplified by passing to the 3-dimensional Einstein
frame metric $g_{\mu\nu} = e^{-4\Phi}G_{\mu\nu}$, in terms of which
the action is
\begin{equation}\label{action}
S = \frac1{2\kappa_3^2}\int d^3\!x\sqrt{-g}
\left( R - 4g^{\mu\nu}\partial_\mu\Phi\partial_\nu\Phi\right).
\end{equation}
Here and below the Ricci scalar and tensor are those derived from $g_{\mu\nu}$. The system we are considering thus consists of a scalar minimally coupled to $2+1$ gravity.\footnote{The analysis that follows would require only a trivial modification for a situation with multiple minimally coupled scalars, for example if one wanted to consider a more complicated decay pathway in which some moduli of the internal seven dimensions were active.}

However, in the system under consideration the supergravity equations of motion will be violated by stringy effects in two different ways. First, the initial orbifold configuration has a conical singularity (with a quantized deficit angle). (If the decay is appropriately fine-tuned, then the final configuration may also have a conical singularity, with a smaller deficit angle.) Second, during the decay itself, the tachyon and (presumably) the massive string fields become active. Furthermore, during this phase the curvatures and energy densities will be large in string units, rendering the supergravity equations of motion invalid. As argued above, however, it is believed that after a sufficiently long time the energy will have spread out over a large region and the massive fields will have dumped their energy into the massless ones.

From a practical point of view, there are two ways we can encode our ignorance about what happens during this stringy phase of the decay. In this section we will mostly treat it as defining some unknown initial conditions, from which we can evolve the supergravity equations of motion forward in time. In the next section, when we write down explicit solutions, it will be more convenient to encode the ``bad" behavior in an unknown source function for the supergravity fields. This function is initially localized at the origin and only sources the metric, but then during the decay sources the dilaton as well (since it must carry away the energy released by the decay), and finally turns off once supergravity becomes valid everywhere.

We choose (for the moment) to work in light-cone coordinates, so that without loss of generality the metric is given by
\begin{equation}\label{metric}
ds^2 = -4e^{2\sigma(u,v)}du\,dv + r^2(u,v)d\theta^2.
\end{equation}
This choice does not fix the gauge completely, as we can still replace
$u$ by an arbitrary function of itself and $v$ by an arbitrary function
of itself and leave the metric in this form.

A simplifying feature of massless scalars is that their stress tensor has the
same structure as the Einstein tensor,
\begin{equation}\label{stress}
\frac{\kappa_3^2}4T_{\mu\nu} = 
\partial_\mu\Phi\partial_\nu\Phi -
\frac12g_{\mu\nu}\partial_\lambda\Phi\partial^\lambda\Phi,
\end{equation}
so that the Einstein equation can be written as
\begin{equation}\label{einstein}
R_{\mu\nu} = 4\partial_\mu\Phi\partial_\nu\Phi.
\end{equation}
With the symmetries we have imposed, this equation has four
non-trivial components. We begin by considering the angular component:
\begin{equation}
R_{\theta\theta} = 4\left(\partial_\theta\Phi\right)^2 = 0.
\end{equation}
Evaluating the Ricci tensor on the metric \ref{metric}, we find that
this implies
\begin{equation}
\partial_u\partial_vr = 0.
\end{equation}
Hence the function $r$ can be decomposed into ``left-moving'' and
``right-moving'' pieces:
\begin{equation}
r(u,v) = r_u(u) + r_v(v).
\end{equation}
Let us assume without loss of generality that both $u$ and $v$ are
future-directed coordinates. Barring the existence of trapped or
anti-trapped surfaces,\footnote{The subsequent analysis can be used to derive a non-singular evolution from any Cauchy surface that does not contain a trapped surface. This shows that a trapped surface cannot form, as expected in $2+1$ gravity.} we can further assume without loss of generality
that $u$ is everywhere outward-directed and $v$ is everywhere
inward-directed, i.e.\ $r_u$ is monotone increasing and $r_v$ is
monotone decreasing. We can therefore use our remaining coordinate
freedom to choose the $u$ and $v$ coordinates such that
\begin{equation}
u = r_u, \qquad v = -r_v.
\end{equation}
The metric is now
\begin{eqnarray}\label{fixed}
ds^2 &=& -4e^{2\sigma(u,v)}du\,dv + (u-v)^2d\theta^2 \nonumber \\
&=& e^{2\sigma(t,r)}(-dt^2+dr^2) + r^2d\theta^2
\end{eqnarray}
(where $t=u+v$). The static orbifold is described by a constant value
of $\sigma$:
\begin{equation}
\sigma = \ln n \qquad (C/Z_n).
\end{equation}
The only remaining coordinate freedom is an overall shift of $t$,
which we will fix (up to an ambiguity of a few string lengths) by saying that the stringy phase of the tachyon condensation occurs at $t=0$.
The three remaining Einstein equations \ref{einstein} in this
coordinate system are
\begin{eqnarray}\label{sigmaeomuv}
R_{uu} = \frac{2\partial_u\sigma}{u-v} &=& 4(\partial_u\Phi)^2 \nonumber \\
R_{vv} = -\frac{2\partial_v\sigma}{u-v} &=& 4(\partial_v\Phi)^2 \nonumber \\
R_{uv} = -2\partial_u\partial_v\sigma &=& 4\partial_u\Phi\partial_v\Phi.
\end{eqnarray}
The consistency of these equations is guaranteed by the conservation
of the stress tensor \ref{stress}. Given $\Phi(u,v)$, they have a
unique solution for $\sigma$ once an initial or final value is
chosen. In $t,r$ coordinates they read:
\begin{equation}\label{sigmaeomtr}
\partial_t\sigma = 4r\partial_t\Phi\partial_r\Phi, \qquad
\partial_r\sigma =
2r\left((\partial_t\Phi)^2+(\partial_r\Phi)^2\right).
\end{equation}

Finally, we have the dilaton equation of motion, $\Box\Phi=0$.
The great advantage of the coordinate system \ref{fixed} is that
$\sigma$ drops out of this equation (assuming again that $\Phi$
doesn't depend on $\theta$); as in flat space, we have
\begin{equation}\label{scalars}
\left(\partial_t^2 - \frac1r\partial_rr\partial_r\right)\Phi = 0.
\end{equation}

The problem thus splits into two steps: first solve the equation of
motion \ref{scalars} for the dilaton; then, to find the gravitational
back-reaction, plug the solution into \ref{sigmaeomuv} or
\ref{sigmaeomtr} and solve for $\sigma$.

\section{Solutions}

As explained at the beginning of section 2,
the supergravity equations of motion are violated during
the stringy phase of tachyon condensation, when $\Phi$
is sourced so that it can carry away the
energy released in the decay. Thus, rather than \ref{scalars}, we actually have
\begin{equation}\label{sources}
\left(\partial_t^2 - \frac1r\partial_rr\partial_r\right)\Phi = 
2\pi\rho(t,r),
\end{equation}
where the source $\rho$ is presumably supported within a few
string lengths of the origin, both in space and in time. Assuming that
$\Phi$ is initially constant with no incoming radiation,
$\Phi(u,\,v\!=\!-\infty)=\Phi^{(0)}$, equation \ref{sources} is in
principle straightforwardly solved using the retarded Green function,
\begin{equation}\label{Green}
G(t,r) = \left\{
\begin{array}{cc}
0,&t<r \\
\frac1{\sqrt{t^2-r^2}},\quad &t>r
\end{array}
\right..
\end{equation}

At late times the leading behavior will be governed by the total
amount of dilaton sourced, $R=2\pi\int dtdr\,r\rho(t,r)$:
\begin{equation}\label{latetimes}
\Phi(t,r) \sim \Phi^{(0)} +
\frac R{\sqrt{t^2-r^2}}, \qquad t-r\gg l_{\rm s}
\end{equation}
(unless $R$ vanishes, in which case the leading behavior will be
governed by the lowest non-zero moment of $\rho$). Note in
particular that in the final static solution ($t\to\infty$ with $r$
fixed) the dilaton returns to its initial value $\Phi^{(0)}$.  If
the order of the final-state orbifold is $n_{\rm f}$, then, using
equations \ref{latetimes} and \ref{sigmaeomtr}, at late times the
geometry will be given by
\begin{equation}
\sigma(t,r) = 
\ln n_{\rm f} - 4r\int_t^\infty dt'\,\partial_t\Phi\partial_r\Phi
\sim \ln n_{\rm f} + \frac{R^2r^2}{(t^2-r^2)^2},
\qquad t-r\gg l_{\rm s}.
\end{equation}

As an example, figure 1 shows a numerical solution to equations
\ref{sources} and \ref{sigmaeomtr} (obtained using {\it Mathematica})
for the decay of $C/Z_3$ to the plane. The source function was
\begin{equation}\label{exsource}
\rho(t,r) = \alpha e^{-t^2-r^2},
\end{equation}
where the normalization $\alpha=0.19$ was set by requiring $\sigma$
to decrease by $\ln 3$.
\EPSFIGURE{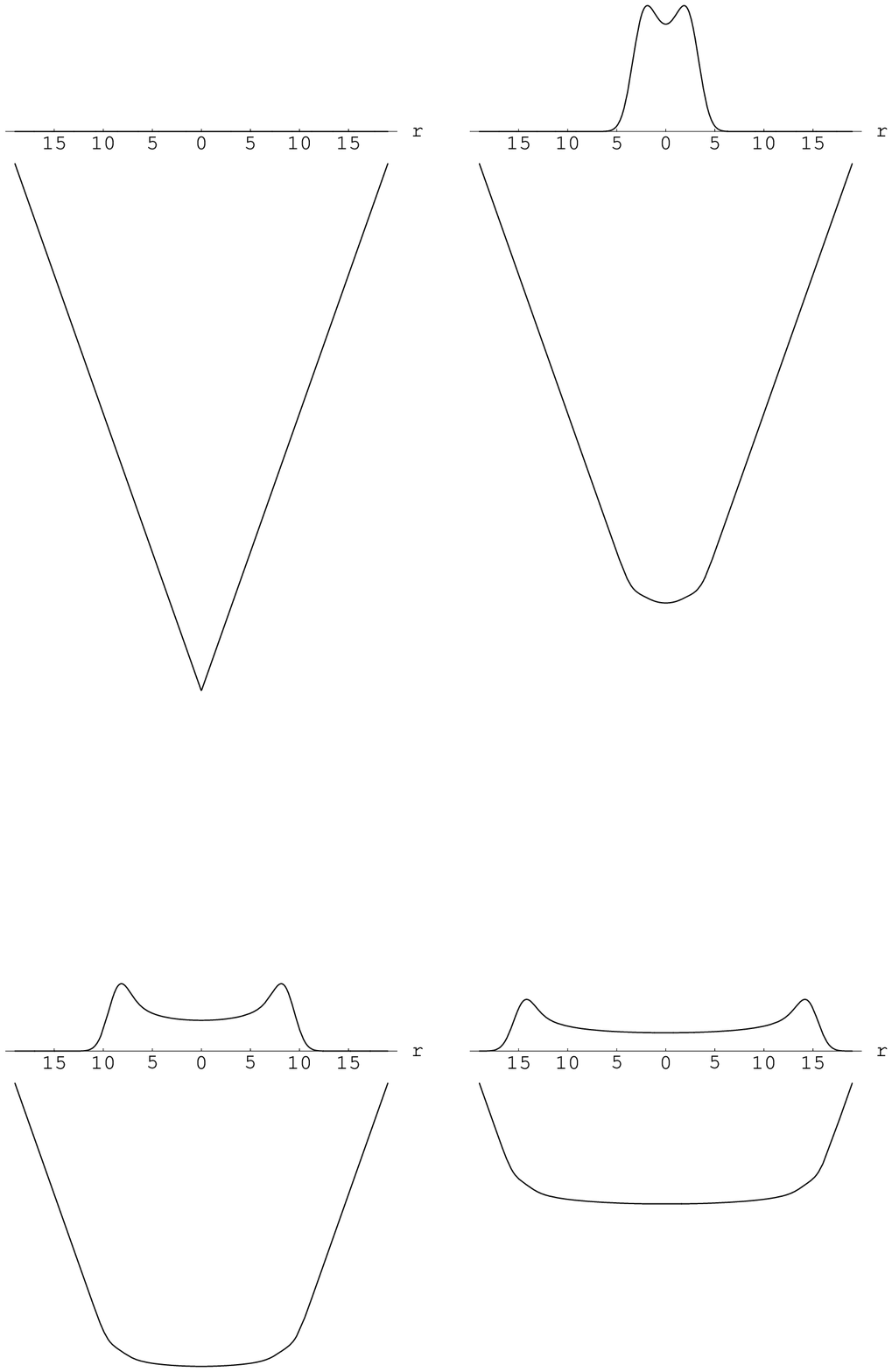,width=4.2in}{Numerical simulation of decay of
$C/Z_3$ to the plane, using the source function \ref{exsource}. Each
of the four figures shows a constant $t$ slice, at $t=-3$ (top left),
$t=3$ (top right), $t=9$ (bottom left), and $t=15$ (bottom right). For
each $t$ the field configuration $\Phi(t,r)$ is plotted above a cross
section of the geometry.}

The change in the order of the orbifold group, from $n_{\rm i}$ to
$n_{\rm f}$, clearly dictates the total amount of energy released in
the form of dilaton radiation. In the limit where the sources are
spatially localized,\footnote{Note that this is a formal limit; with a point source for
$\Phi$ the energy
density diverges like $1/r^2$ near the origin, leading to a
geometrical singularity as $\sigma$ goes to $-\infty$ (like $\ln r$)
there.}
\begin{equation}
\rho(t,r) = \delta^2(r)\rho(t),
\end{equation}
we can see this connection explicitly in the following relation,
which can be derived from equations \ref{Green} and \ref{sigmaeomtr}:
\begin{equation}\label{deltasigma}
\ln \frac{n_{\rm f}}{n_{\rm i}} = \Delta\sigma 
=-2\int_0^\infty d\omega\,\omega|\tilde\rho(\omega)|^2,
\end{equation}
where $\tilde\rho(\omega)=\int dt\,e^{-i\omega t}\rho(t)$. Equation \ref{deltasigma} establishes that $n_{\rm f}\le n_{\rm i}$; this is a special case of the theorem that Bondi energy
decreases as a function of light-cone time. Note that $\Delta\sigma$
is independent of $R=\tilde\rho(0)$.

It is possible to take a further limit where the source is
localized in time as well, holding $\Delta\sigma$ fixed. In the
resulting solution, $\Phi$ is constant, while $\sigma$ is a step
function of $v$:\footnote{To be slightly more precise, this solution
can be obtained by considering the following field
configuration:
\begin{eqnarray*}
\sigma &=& \ln n_{\rm i} + \Delta\sigma\,f(v/\epsilon) \\
\Phi &=& \Phi^{(0)} + \sqrt{-\Delta\sigma\,\epsilon/2u}f(v/\epsilon),
\end{eqnarray*}
where
$$
f(x) = \left\{\begin{array}{cc}
0,\quad&x\le0 \\ x,\quad & 0\le x\le1 \\ 1,\quad & x\ge1 
\end{array}\right..
$$
At finite $\epsilon$ this configuration does not quite satisfy the
equations of motion, but the errors vanish in the limit
$\epsilon\to0$.}
\begin{equation}
\sigma(u,v) = 
\left\{
\begin{array}{cc}
\ln n_{\rm i},\quad & v<0 \\
\ln n_{\rm f},\quad & v>0
\end{array}
\right..
\end{equation}

\acknowledgments
I would like to thank A. Adams, A. Dabholkar, A. Fayyazuddin, R. Gopakumar, H. Liu, G. Mandal, S. Minwalla, J. Raeymaekers, and A. Sen for useful discussions, and S. Minwalla for useful comments on the manuscript. I would especially like to thank A. Fayyazuddin for suggesting this problem to me while we were both visiting the Perimeter Institute. This work was completed while I was a Visiting Fellow at the Tata Institute for Fundamental Research. I would also like to thank the Harish-Chandra Research Institute for hospitality while some of this work was complete. I am supported by a Pappalardo Fellowship from MIT. This work is supported in part by funds provided by the U.S.\ Department of Energy under cooperative research agreement DF-FC02-94ER40818.

\end{document}